\documentclass{statsoc}

\usepackage[a4paper]{geometry}
\usepackage{graphicx}
\usepackage[textwidth=8em,textsize=small]{todonotes}
\usepackage{amsmath}
\usepackage{natbib}
\usepackage{algorithm}
\usepackage[noend]{algpseudocode}
\usepackage{multicol}
\usepackage{amssymb}
\usepackage{stmaryrd}
\usepackage{longtable}

\title[Wavelet Screaming: a novel approach to analyzing GWAS data]{Wavelet Screaming: a novel approach to analyzing GWAS data}
\author[William Denault et al.]{William Denault \thanks{Corresponding author.}}

\address{Department of Genetics and Bioinformatics, Norwegian Institute of Public Health, Oslo;
\\ Department of Global Public Health and Primary Care, University of Bergen, Bergen, Norway}
\email{william.denault@gmail.com}

\author{H\r{a}kon K. Gjessing}
\address{Centre for Fertility and Health (CeFH), Norwegian Institute of Public Health, Oslo, Norway; Department of Global Public Health and Primary Care, University of Bergen, Bergen, Norway}

\author{Julius Juodakis}
\address{ Department of Genetics and Bioinformatics, Norwegian Institute of Public Health, Oslo, Norway; Department
of Obstetrics and
Gynecology, Institute of Clinical Sciences, Sahlgrenska Academy , University of Gothenburg, Gothenburg, Sweden}

\author{Bo Jacobsson}
\address{Department of Genetics and Bioinformatics, Norwegian Institute of Public Health, Oslo, Norway; Department of Obstetrics and Gynecology, Institute of Clinical Sciences,Sahlgrenska Academy , University of Gothenburg, Gothenburg, Sweden }

\author[William Denault et al.]{Astanand Jugessur}
\address{ Department of Genetics and Bioinformatics, Norwegian Institute of Public Health, Oslo, Norway; Centre for Fertility and Health (CeFH), Norwegian Institute of Public Health, Oslo, Norway; Department of Global Public Health and Primary Care, University of Bergen, Bergen, Norway}

\begin{document}

\begin{abstract}

We present an alternative method for genome-wide association studies (GWAS) that is more powerful than the regular GWAS method for locus detection. The regular GWAS method suffers from a substantial multiple-testing burden because of the millions of single nucleotide polymorphisms (SNPs) being tested simultaneously. Furthermore, it does not consider the functional genetic effect on the response variable; i.e., it ignores more complex joint effects of nearby SNPs within a region. Our proposed method screens the entire genome for associations using a sequential sliding-window approach based on wavelets. A sequence of SNPs represents a genetic signal, and for every screened region, we transform the genetic signal into the wavelet space.  We then estimate the proportion of wavelet coefficients associated with the phenotype at different scales. The significance of a region is assessed via simulations, taking advantage of a recent result on Bayes factor distributions. Our new approach reduces the number of independent tests to be performed. Moreover, we show via simulations that the Wavelet Screaming method provides a substantial gain in power compared to the classic GWAS modeling when faced with more complex signals than just single-SNP associations. To demonstrate feasibility, we re-analyze data from the large Norwegian HARVEST cohort.
\\\\
\textit{Keywords}: Bayes factors, GWAS, SNP, Multiple testing, Polygenic associations, Wavelets.

\end{abstract}

\section{Introduction}

The objective of a genetic association study is to identify the location of genetic regions (loci) that are associated with a phenotype of interest. Although the human genome is very similar across individuals, it is interspersed by single base-pair differences (SNPs) that contribute to the observed differences across individuals. One of the most common approaches to uncovering genetic associations for a given trait or disorder is to conduct a genome-wide screening for associations (GWAS) where the significance of the effect of each SNP on a phenotype of interest is assessed in a sequential fashion. Despite its many successes, this approach is limited by two important issues: (i) it incurs a substantial multiple-testing burden, and (ii) it ignores the functional nature of the genetic effect by failing to exploit the dense genotyping of the genome.  Because the genome is a code for the observable phenotype and only considers the change of one incremental unit \textit{per se} (i.e., one additional variant), it is unlikely that a SNP alone would be able to efficiently model how a change in the genome might mirror a change in the phenotype. 

To address these issues, we developed the Wavelet Screaming approach by harnessing insights from functional mixed modeling  \citep{MR06}. More specifically, we adopt the approach described by \cite{SS15} where the authors first tested for association with a functional phenotype (the response signal) by transforming the signal using fast discrete wavelet transform \citep{M98} and then performing single-SNP association tests with the spectrum. In essence, our main idea is to reverse this approach. The use of reverse regression to identify genetic loci is now more widespread in the genetic literature \citep{H17}. Our approach entails treating sizable chunks of the genome ($\approx 1$ million base pairs) as the functional phenotype. We then essentially perform a dimensional reduction using wavelet transform and subsequently test for associations between the wavelet coefficients and the true phenotype (considered here as an endogenous variable).  
 
The issue of multiple testing can be resolved using other regularization methods, such as the Fused Lasso \citep{T05}. The principle of Fused Lasso is to perform a penalized regression that takes into account how variables (i.e., SNPs) that are physically close to each other might have similar effects. Fused Lasso can then define a region of association between the SNPs and the phenotype. However, Fused Lasso performs local testing, whereas it is preferable to conduct a broader test by estimating the fraction of wavelets (blocks) associated at each scale. Such broader testing combined with multiple levels of information may provide additional insights into the mechanism underlying the genetic association.

Despite an increased interest for penalized regressions in the statistical community, they seldom appear in the top-tiered genetics publications. Although penalized regression has recently been added to one of the leading software for GWAS --
 PLINK \citep{PN07}, the lack of a comparable software for meta-analysis is a major limitation of this approach. This is because a comprehensive genome-wide association meta-analysis (GWAMA) typically relies on summary statistics from multiple cohorts. Even if meta-analyses are now doable in the Lasso regression setting \citep{LT14}, they are not currently available for variants of Lasso regression or for other regularization penalties. By contrast, our method is easily amenable to meta-analysis through the use of the classic Fisher \citep{Fish} method to combine p-values for loci that show associations in different cohorts.

One of the main drawbacks of a straightforward wavelet regression is that the considered sets of SNPs are not of length $2^J$ and the physical locations of the SNPs are not evenly spaced. To handle this, we suggest using the robust wavelet regression developed by  Kovac and Silverman \citep{KS00} .  
By reversing the regression and targeting a given region for association, we can tackle these two main issues by: 
\begin{enumerate}
\item Using regional association instead of SNP-specific association to reduce the number of tests to be performed from 8 million (for common SNPs) to approximately 6000 by using overlapping loci of 1 Mbp in length.
\item Using a robust wavelet regression for non-equally spaced data that takes into account the functional nature of the genetic effect, i.e., more complex joint effects of multiple nearby SNPs within a region.
\end{enumerate}
The remainder of our paper is structured as follows. We first describe the statistical setting and the wavelet methodology used to generate the wavelet coefficients. Next, we describe how to compute the likelihood of the association between the wavelet spectrum and the phenotype $\Phi$. The phenotype $\Phi$ in this paper is a univariate vector of a continuous or a binary trait. After a comprehensive evaluation of the method using different simulations, we apply it to the Norwegian HARVEST dataset, which is a subproject of the Norwegian Mother and Child Cohort Study (MoBa), \cite{Harvest}) with a special focus on children's gestational age at birth.

\section{Description of Wavelet Screaming}

\subsection{Context}

\subsubsection*{Wavelet representation of the genome}

We first process the multi-SNP data using a wavelet transform.  At first sight it might seem odd to transform the raw data into a different format. However, this type of approaches is widely used in the ``Gene- or Region-Based Aggregation Tests of Multiple Variants" (\cite{LG14}). In these methods, like in the Burden test, the effects of the genetic variants in a given region are summed up to construct a genetic score that can then be used in the regression. \\

In the following section, we assume some familiarity with wavelet transform. A comprehensive introduction to wavelets is available in \cite{WM}. In the rest of this article, wavelet `(transform)' specifically refers to the `Haar wavelet (transform)'.
We code a SNP $0$ if an individual is homozygous for the reference allele, 1 if heterozygous, and 2 if homozygous for the alternate allele -- consistent with an additive genetic model \citep{PN07}. Although this coding is arbitrary, it is the standard way of coding \citep{PN07}. Our procedure is resilient to this coding  (see Section 3).
\\\\
For the  scale $0$, the wavelet coefficient $d$ and $c$ can be interpreted the same way:

\begin{itemize}
\item The coefficient at scale $0$ for an individual summarizes the amount of discrepancy between the individual's genotypes and the reference genotypes coded as $0...0$ (which is essentially what is tested in gene/regional tests).
\end{itemize}

The $d$ wavelet coefficients at scale $s> 0$ can be interpreted as:

\begin{itemize}
\item The wavelet $d$ coefficient at scale $s$ and location $l$  for an individual represents the difference in the number of minor variants between the left part of the region (defined by $s,l$) and the right part.
\end{itemize}

The $c$ wavelet coefficients at scale $s> 0$ can be interpreted as:

\begin{itemize}
\item The wavelet $c$ coefficient at scale $s$ and location $l$ for an individual represent the amount of discrepancy between the individual's genotypes and the reference genotypes coded as $0...0$ for the region defined by $s,l$. 
\end{itemize}

The main rationale behind this modeling is that, if there is a causal effect of an allele on the phenotype, the association is likely to be spread across genomic regions of a given size (scale) at different positions (locations). By using the wavelet transform to perform a position/size (time/frequency) decomposition and then regressing on the wavelet coefficients, we are able to visualize \textit{where} (location) and \textit{how} (scale) the genetic signal influences the phenotype. 
\\

In the rest of this article, we use `wavelet coefficients' to specifically refer to any one of two coefficients, $c$ or $d$, but never to both at the same time. In Section 5 we discuss the use of $d$ or $c$ coefficient.  In the rest of this section, using $c$ or $d$ coefficients does not change the general framework. Let us define a genetic region of a chromosome $c$ of individual $j$ as the set of SNPs $G$. $G$ has physical positions (base pair, $bp$) between a lower bound $lb$ and an upper bound $ub$. It follows that a given genetic region for individual $j$ can be written as:
\begin{equation}GR_{lb,ub,j} = \{ G_{bp,j} , lb < bp < ub  \} \end{equation}

\begin{figure}[!ht]
\begin{center}
\includegraphics[scale=0.5]{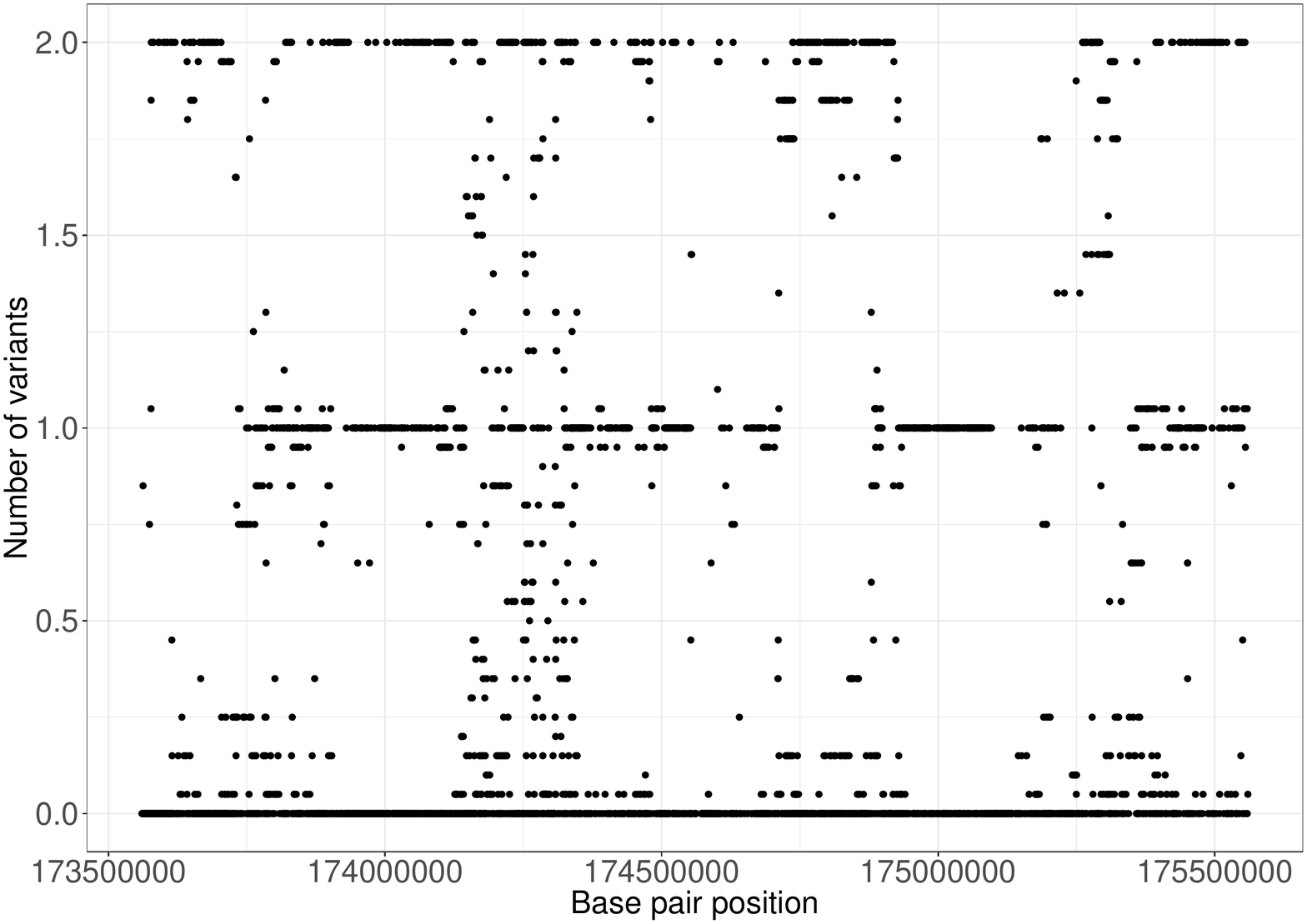}
\caption{Genetic variation of one individual within a locus consisting of 2 million base pairs (including 10000 imputed SNPs).}

  \label{fig:loci}
  \end{center}
\end{figure}

We consider $GR_{c,lb,ub,j}$ as a function/signal and observe this signal for every individual at pre-determined increasing positions $bp_1,..., bp_n$, with some error due to the genome-wide imputation process \citep{LW09}. Thus,

\begin{align}
G_{bp,j} & = GR_{lb,ub,j}(bp)+ \epsilon_{bp,j}\\
& \epsilon_{bp}  \quad  \text{iid}\quad  N(0,\sigma^2_{bp}) \nonumber 
\end{align}
where $G_{bp_i,j}$ is the measured genetic signal (SNP) at position $bp$ for the individual $j$. The variance $\sigma^2_{bp}$ can be interpreted as a function of the imputation quality $IQ$ which has a value in $\left[0,1\right]$.  $1$ represents a perfectly imputed SNP or genotyped SNP; thus, $\sigma^2_{bp} \propto 1-IQ_{bp}$. As the data are preprocessed by having undergone specific criteria for quality control, only SNPs with an  $IQ\in \left[0.7,1\right]$ are retained in further analysis. We assume that the imputation qualities are independent and heteroscedastic. 
As the value of a SNP is in $\llbracket 0,2\rrbracket$ and then in $\left[ 0,2\right]$ in the dosage convention after imputation \citep{PN07}, the hypothesis of error normality is arguable. In future developments, we might consider a more suitable noise law.

\subsection{Preprocessing}
\subsubsection*{Non-decimated wavelet transform}
We use the method of \cite{KS00} for non-decimated and unevenly spaced data. This method takes an irregular grid of data (e.g., representing the sampling of the different genetic regions) and interpolates the missing data into a pre-specified regular grid. More precisely, for a genetic region $GR_{lb,ub}$ measured at positions $ bp_1 ... bp_n $, a new grid of points is defined as 
$ t_{0},...,t_{N-1}$, where $N=2^J, J\in \mathbb{N} $, $t_{k}= (k+\frac{1}{2})2^{-J}$ and $J = min\{j \in \mathbb{Z}, 2^j \geq n \}$. We interpolate the signal on this grid and run the classic wavelet transform to obtain the wavelet coefficients. In practice (see Section $4$), we recommend selecting  genetic regions that have a relatively high density of imputed SNPs.

\subsubsection*{Coefficient-dependent thresholding and quantile transform}

For each individual wavelet decomposition, we use the VisuShrink approach \citep{KS00} to shrink the interpolated wavelet coefficients and reduce the dependence between the wavelet coefficients within scales. This method estimates the variance of each wavelet coefficient before determining a specific threshold for each wavelet coefficient. By determining coefficient-dependent thresholds using the wavelet coefficient variance, we can account for the individual heteroscedasticity of the noise.  We then quantile-transform each wavelet coefficient distribution within the population. This ensures that the endogenous variables used in the regressions are normally distributed. Particularly in cases where there were no associations, the residuals were normally distributed.

\subsection{Modeling}
To gauge the effect of a genetic region on the phenotype, we first need to assess whether certain scales are associated with the phenotype at different locations. Let $\pi$ be a vector of length $J$ where $ \forall j \in [0:J], \pi_j \in [0,1] $, where $\pi_j $ represents the proportion of wavelet coefficients at scale $j$ associated with the phenotype. To assess the significance of a genetic region, we want to test the following hypothesis:

\begin{equation}
H_0:\pi =(0,...,0) \enspace vs \enspace H_1:\exists j \in [0:J], \pi_j \ne 0
\end{equation}

Below, we describe our test statistic (likelihood ratio) and how we compute its different components and test its significance.

\subsubsection{Bayes factors}

To test for association between the phenotype and the wavelet coefficient $\tilde{G}_{sl}$ for a given genetic region, we perform a regression between the wavelet coefficient and the phenotype $\Phi$ using Normal-Inverse-Gamma (NIG) prior. It is important to correct for confounding factors $C$ in a GWAS. Through our framework, we can readily incorporate those confounding factors into the regression models.

The association models for each scale and location are defined as follows: 
\begin{align}
&M_0: \tilde{G}_{sl} = \beta_{sl,0} +\beta_{sl,C}C+\epsilon \nonumber \\
&M_1: \tilde{G}_{sl} = \beta_{sl,0} + \beta_{sl,1}\Phi +\beta_{sl,C}C+\epsilon 
\end{align}
where $C$ is a matrix of dimension $c \times 1$ and $\beta_{sl,C}$ is a matrix of dimension $1 \times c$. We compute the Bayes factors of the wavelet regression $sl$ using the closed form provided by \cite{SS07} for NIG prior with  $\sigma=0.2$.

\subsubsection{Ratio statistic}

Our goal is to assess the significance of the vector $\pi = (\pi_0,...,\pi_s)$ where $\pi_s$ represents the proportion of wavelet coefficients at scale $s$ that is associated with the phenotype $\Phi$, and $\tilde{G}$ is the wavelet representation of the genotype. To test the significance of $\pi$, we construct a test statistic by computing the following likelihood ratio:

\begin{equation}\label{eq:1}
\Lambda (\pi , \tilde{G} ,  \Phi ) = \frac{p( \tilde{G} |\pi ,  \Phi) }{p( \tilde{G} |\pi\equiv 0 ,  \Phi)} 
\end{equation}

Following the method of \citet{SS15}, we denote $\gamma_{sl}$ as the random variable with support $\{0,1\}$. $\gamma_{sl}=1$ when the wavelet coefficient $\tilde{G}_{sl}$ is associated with the variable $\Phi$; $0$ when there is no association. We consider $\pi$ as a hyperparameter of $\gamma_{sl}$; i.e., 

\begin{equation}
p(\gamma_{sl} =1| \pi) = \pi_s
\end{equation}

\cite{SS15} assume independence between the wavelet coefficients. However, this is unlikely to hold in practice because of the correlation structure of the genetic regions. Under the assumption of independence of the wavelet coefficients, we can rewrite \ref{eq:1} as follows:

\begin{align}
\Lambda (\pi , \tilde{G} ,  \Phi ) & = \prod_{s,l} \frac{p( \tilde{G}_{sl} |\pi_s ,  \Phi) }{p( \tilde{G}_{sl} |\pi_s = 0 ,  \Phi)} \\
& = \prod_{s,l} \frac{\pi_s p( \tilde{G} |\gamma_{sl}=1  ,  \Phi) + (1-\pi_s) p( \tilde{G} |\gamma_{sl}=0  ,  \Phi) }{p( \tilde{G} |\gamma_{sl}=0 ,  \Phi)} \label{prod like}
\end{align}

We denote $BF_{sl} (\tilde{G},\Phi) = \frac{p( \tilde{G}_{sl} |\gamma_{sl}=1,  \Phi) }{p( \tilde{G}_{sl} |\gamma_{sl}= 0 ,  \Phi)} $ as the Bayes factor of the association between the wavelet coefficient at scale $s$ and location $l$. Computation of the Bayes factor will be discussed below. Using this notation, we can rewrite \ref{prod like} as:

\begin{equation} \label{eq:Lambda}
\Lambda (\pi , \tilde{G} ,  \Phi ) = \prod_{s,l} \left[ \pi_s BF_{sl}  +(1-\pi_s)\right]
\end{equation}

We can then compute the likelihood ratio statistic by maximizing the lambda statistics over $\pi $ and determining $\pi$ via the EM algorithm. 
\begin{equation}
\hat{\Lambda} (\tilde{G} ,  \Phi ) = max_{ \pi \in [0,1]^s}   \Lambda (\pi , \tilde{G} ,  \Phi ) 
\end{equation}

\subsubsection{Significance of a genetic region}

As the distribution of $\Lambda$ is unknown, we simulate $\Lambda$ under $H_0$ by simulating $BF_{sl}$ under the hypothesis of no association. Recently, \cite{ZG17} showed that under $H_0$ and a wide spectrum of priors, the Bayes factors (including the NIG prior) for a Gaussian model follow a specific law. More precisely,

\begin{equation}
2log(BF) = \lambda_1 Q_1+ log(1-\lambda_1)
\end{equation}

where $Q_1$ is a non-central chi-squared random variable with $df=1$, and $\lambda_1$ and its non-centrality parameter have a closed form. Using reverse regression, $\lambda_1$ can be computed directly from the design matrix for all the loci at once. Even though the non-centrality parameter is dependent on the wavelet coefficients, it vanishes to zero asymptotically with increasing sample size.

Indeed, \cite{ZG17} showed that for $df=1$ Bayes factors, $Q_1$ is asymptotically equal to the likelihood ratio test statistic for Gaussian linear models. In other words, it is equal to a simple chi-squared statistic with 1 degree of freedom. For large sample sizes (e.g., those typical in a GWAS setting), the non-centrality parameter can be safely set to zero. 
We can then perform $M$ independent simulations of the vector of Bayes Factor under $H_0$. Then, for each simulation $m$, we compute $\hat{\Lambda}_{ m } = max_{ \pi \in [0,1]^s}  \Lambda (\pi ,BF_m) $ using the procedure described above.

\subsubsection*{P-value as a tail estimation problem}
One may suggest computing the p-value via the classic method from the permutation procedure. However, users of this procedure often fail to check the number of permutation needed to obtain reliable p-values, especially at the low end of the scale. By using the normal approximation of the estimation \citep{PML}, the number of permutations $k$ required to obtain a reliable p-value has to be more than $\frac{1}{4P^2}$, where $P$ is the desired level of significance.
However, in our case, we need reliable p-value at the level of significance $\frac{0.05}{6000}\approx 8\times10^{-6}$ for a full genome screen, which would imply $\approx 4 \times 10^{10}$  simulations.
To avoid this computational burden, we suggest using a smaller number of simulations ($10^7$) to estimate the tail distribution of $\hat{\Lambda}$ by fitting a Generalized Pareto distribution. \cite{EVTP} provided an extensive review on the use of the tail approximation by Generalized Pareto distribution to assess the significance of tests. We assume that the tail distribution $\hat{\Lambda}$ can be modeled by extreme-value distributions (i.e., we assume $\hat{\Lambda}$ is in the domain of attraction of one of the extreme-value distributions). 
 Estimation of the high quantile by Generalized Pareto distribution fitted by maximum likelihood leads to correct inference (\cite{HW87}), especially when the estimated quantile is within the sample. We perform the fitting using a threshold determined by the Van Kerm’s rule of thumb \cite{VANKE}. Next, we use this fitted distribution to estimate the associated quantiles of the observation, and thus the  p-values.

\section{Simulations}

\subsection{Complex genetic signals}

We performed simulations for a complex genetic signal by combining real genetic data with a simulated phenotype. We used the locus displayed in Figure \ref{fig:loci}, computed the linkage disequilibrium (LD) structure, and selected $28$ SNPs located at the center of each LD block. We performed two sets of simulation:
\begin{itemize}
\item \textit{Mono-directional}:  for each iteration, we randomly selected $1$ to $28$ SNPs. For each individual, we summed their SNP dosages within the selected set of SNPs to construct a score. On top of the individual score, we added normally distributed noise, scaled so that the genetic score explains $0.5\%$ of the total phenotypic variance.

\item \textit{Random direction}: the same setting as above, but the sign of the effect (positive/negative) for each SNP is taken at random. Compared to the mono-directional simulation, where any additional variant raises the level of the phenotype, this is not necessarily the case for random direction. These simulations are made to showcase the sensitivity of Wavelet Screaming to the direction of the SNP coding.

\end{itemize}

\begin{figure}[!ht]
\begin{center}
\includegraphics[scale=0.4]{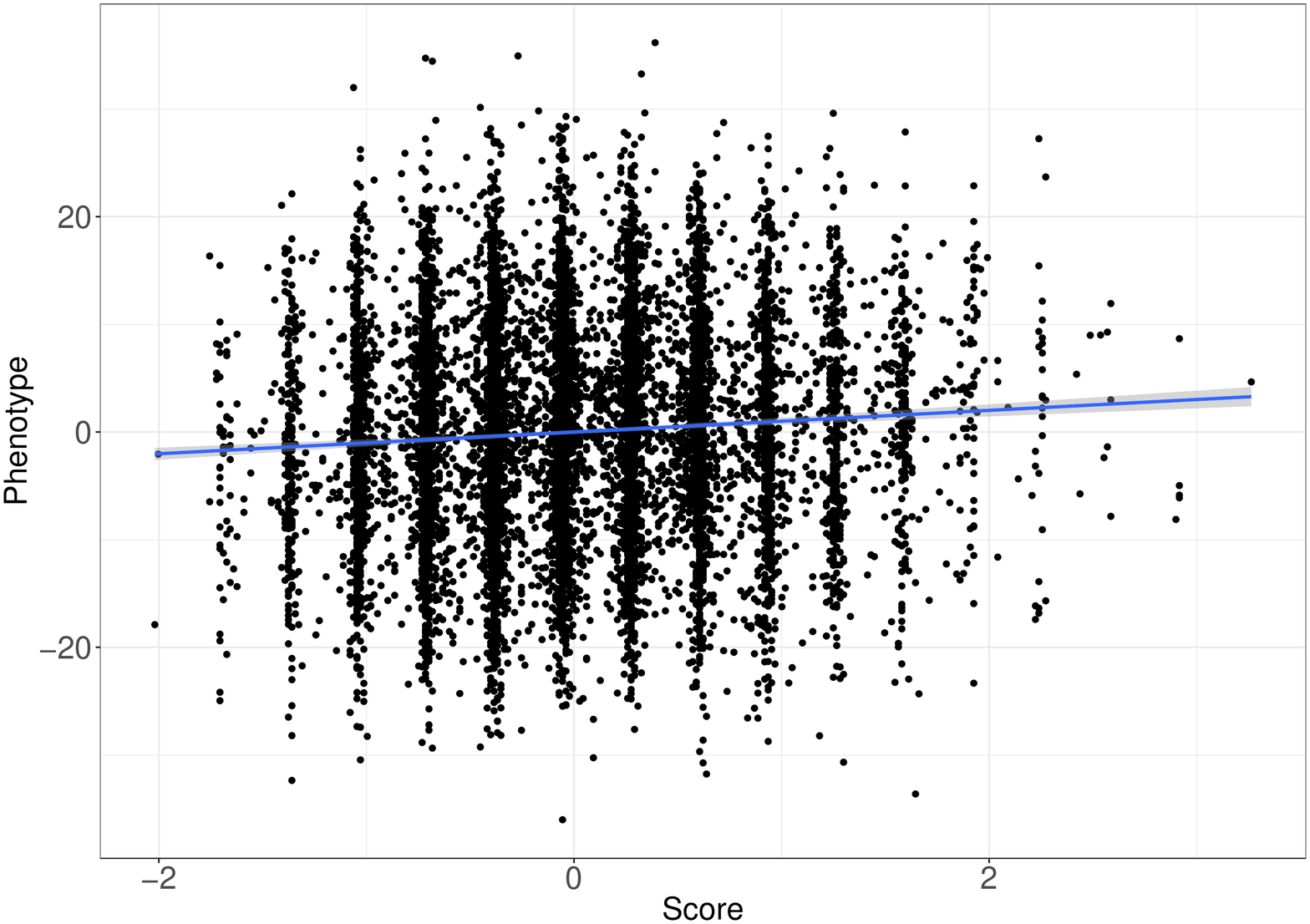}
\caption{Simulated phenotype against the generated score (20 SNPs selected) }
  \label{fig:simu}
  \end{center}
\end{figure}

These simulations are engineered to mimic a diluted effect of SNPs within different LD blocks, or ``block polygenic effect'' where each variant has a small additive effect in the same direction. The variance explained by a single SNP varies between $0.5\%$, which is typical for the top SNPs in a GWAS (\cite{FM14}), to $0.018\%$, at which level variants are normally not detected by the standard GWAS framework.

We performed Wavelet screaming on these simulations using $c$ and $d$ coefficients. Because the design matrices do not vary between iterations (the phenotype varies but not the genotype), we can compute $\lambda_1$ directly using its closed form. We obtained $\lambda_1 = 0.99974$. As the average of the simulated phenotype increases with decreasing $\lambda_1$ (at least near $1$), to be conservative, we performed the simulations of $\hat{\Lambda}$ using $\lambda_1= 0.9997$. In total, we performed $10^6$ simulations of $\hat{\Lambda}$. \\

\begin{figure}[!ht]
\begin{center}
\includegraphics[scale=0.4]{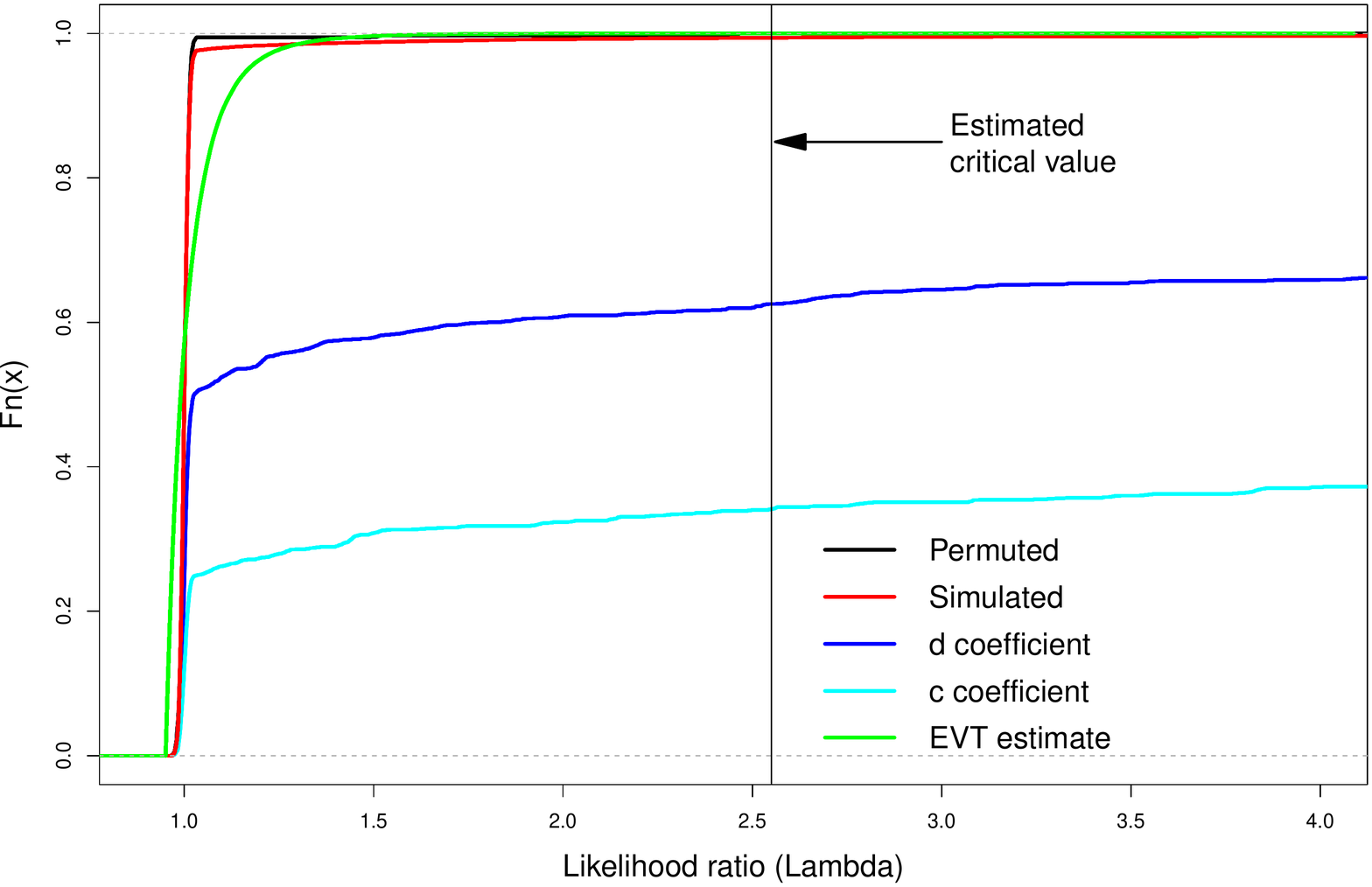}
\caption{Empirical cumulative distribution functions from the simulations (c and d coefficients) corresponding to the $\hat{\Lambda}$ for mono-directional simulations.}

  \label{fig:ECDF}
  \end{center}
\end{figure}

We computed $\hat{\Lambda}$ on $1000$ index permutations between the scores and the genotype to obtain the null distribution of $\hat{\Lambda}$. The simulated distribution has slightly heavier tails than the permuted dataset (see Figure \ref{fig:ECDF}). Therefore, p-values computed from the simulated distribution for large values of $\hat{\Lambda}$ will tend to be conservative. We further computed the p-values using the fitted distribution of the simulated $\hat{\Lambda}$. As the null distributions are 'spiked' around 1, we provide in the Annex a zoomed plot around $1$.

\subsection{Wavelet screaming improves the discovery rate}

Table \ref{table:ConpA} provides a summary of the results of the above simulations.  Compared to the standard GWAS procedure, Wavelet Screaming clearly improves discovery rate for both $c$ and $d$ coefficients. To investigate this further, we checked the dependency of power and number of components within the signal. These results are displayed in Table \ref{table:ConpB}.

Using a linear model, Wavelet Screaming and the standard GWAS have roughly the same power for a single-SNP effect. However, when there is dilution of an effect due to polygenicity, Wavelet Screaming performed substantially better for both $d$ and $c$ coefficients. Via logistic regression of power on the number of components, we find a significant difference in slopes (Fisher test for general linear model at $p < 10^{-5}$) but not in intercepts.

It appears that for mono-directional Wavelet Screaming with $c$ coefficients, the effect dilution has a positive impact on the discovery rate. This might be due to the simulation setup, as all the SNPs are assumed to have an effect in the same direction. As the $c$ coefficients represent the amount of variation within a region, $c$ coefficients mimic the local score correctly. However, even if the variant is arbitrarily coded as having a positive effect (i.e., random direction simulation), we still obtain a clear improvement in the discovery rate. This demonstrates the robustness of Wavelet Screaming to potential ``misspecification'' in the coding. As far as we know, no other regional-based test can handle this problem without assuming a rare variant distribution.

The results in Table \ref{table:ConpB} also suggest that the dilution effect is highly non-linear for the GWAS linear modeling, with a steep elbow-shaped curve. In contrast, the power for Wavelet Screaming $d$ and $c$ coefficients decreases roughly linearly with the number of components in the score.

\begin{longtable}[]{@{}cccc@{}}

Simulation type & Method & Significance criterion & Power \tabularnewline
\hline
\endhead
Mono-directional & WS $c$ & p-value $\leq 8\times 10^{-6}$ & $67.2\%$\tabularnewline
Mono-directional & WS $d$ & p-value $\leq 8\times 10^{-6}$ & $37.6\%$\tabularnewline
\hline
Random direction & WS $c$ & p-value $\leq 8\times 10^{-6}$ & $46.2\%$\tabularnewline
Random direction &  WS $d$ & p-value $\leq 8\times 10^{-6}$ & $51.2\%$\tabularnewline
\hline
Not applicable & GWAS LM&  p-value $\leq 5\times 10^{-8}$ &  $21.8\%$ \tabularnewline

\caption{Wavelet screaming and GWAS signal detection capacity.}
\label{table:ConpA}
\end{longtable}

\begin{longtable}[]{@{}c cccccc@{}}

Simulation type & Method & $1-5$ & $6-10$ & $11-15$ & $16-20$ & $\geq 21$ \tabularnewline
\hline
\endhead
Mono-direction & WS $c$ & $67 \% $& $50\% $ & $57\% $ & $71\% $  & $75 \%$ \tabularnewline
Mono-direction & WS $d$ & $72 \% $& $48\% $ & $36\% $ & $29\% $  & $19\%$ \tabularnewline
\hline

Random direction & WS $c$ & $64 \% $& $50\% $ & $45\% $ & $37\% $  & $36 \%$ \tabularnewline
Random direction & WS $d$ & $64 \% $& $58\% $ & $55\% $ & $41\% $  & $45\%$ \tabularnewline
\hline

Not applicable & GWAS LM& $74\%$ & $17\%$ & $13\%$ & $9\%$ & $6\%$ \tabularnewline
\caption{Power of the different methods depending on the number of components in the simulation (dilution effect). }
\label{table:ConpB}
\end{longtable}

We conclude that Wavelet Screaming for $d$ and $c$ coefficients does not significantly improve the discovery rate for loci that harbor a single causal SNP, but it is superior than the standard GWAS framework in detecting regions with multiple signals. Thus, in a general setting, Wavelet Screaming for $d$ coefficient is more powerful than standard GWAS. In addition, for SNPs that have an effect pointing in the same direction, Wavelet Screaming with $c$ coefficient provides a substantial improvement in power.

\section{Application}
To test the utility of our new method, we performed a chromosome-wide association study of human gestational duration using Wavelet Screaming. Gestational duration is a complicated phenotype to study in a GWAS setting because of a combination of large measurement errors ($\approx 7$ days,  \cite{ME} ) and typically small genetic effects ($\approx 1.5$ days \citep{obs}). We used GWAS data on mothers from the Norwegian HARVEST study \citep{Harvest} to replicate the lead SNPs reported in the largest GWAS to date on gestational duration \citep{obs}. These SNPs are located on chromosome 5, near the gene for Early B cell factor 1, \textit{EBF1}, and are likely to regulate \textit{EBF1} gene activity. By using the same methodology as in \cite{ZG17}, the lowest p-value obtained in our dataset was $2.8\times 10^{-6}$, which is not considered statistically significant in the classic GWAS setting.

\subsection{Definition of the regions and choice of resolution }

Although millions of SNPs can now be interrogated in a typical GWAS, several chromosomal regions are characterized by poor marker density, in particular near telomeres and centromeres, and in regions of low imputation quality. Most SNPs with low imputation quality are routinely discarded during quality control after imputation. As we preprocess our data using an interpolation, we aim to avoid analyzing purely interpolated regions. Our strategy entails including an additional criterion in the preprocessing step to exclude these types of regions. We propose studying regions of size $1Mbp$ (mega basepairs), with a maximum distance of $10kb$ between any two SNPs. Further, we define overlapping regions where the signals are at the boundary of a given region. By applying these additional criteria, we excluded $18\%$ of the SNPs and defined $253$ regions on chromosome 5.

In addition to avoiding fully interpolated regions, we also need to decide how deep into the wavelet decomposition we would like to analyze. We know that the precision of the wavelet coefficient depends on the amount of non-interpolated points in a given region \citep{KS00}. As a rule of thumb, we propose to have at least $10$ SNPs on average for each wavelet coefficient. Following this preassigned criterion, we ended up with a median spacing between SNP of $202$ base pairs. This means that if we divide each locus of $1Mb$ into $2^9 =512$ subregions, we would on average have $\frac{10^6}{2^9}\times\frac{1}{202} \approx 9.7 $ SNPs per subregion.

\subsection{Model and results }

We applied the Wavelet Screaming approach using the $d$ and $c$ coefficients to the aforementioned dataset on gestational duration. In our modeling, we included the first $6$ principal components for each wavelet regression to control for residual population structure \citep{PCA}. We computed $\lambda_1$ via the analytic formula of \cite{ZG17} and obtained $\lambda_1= 0.9999687$. We simulated $10^7$ $\Lambda$ values under these conditions ($\lambda_1= 0.9999687$, resolution $=9$). We then estimated the parameters of the general Pareto distribution by setting the location parameter at the minimum of the simulated $\Lambda$ value ( $\approx 1$). The shape parameter $\xi = 0.1705$ (confidence interval $(0.1695-0.1715) $ and the scale parameter $\beta = 0.0103$ $(0.01027-0.01033)$).

\begin{figure}
\begin{center}
\includegraphics[scale=0.6]{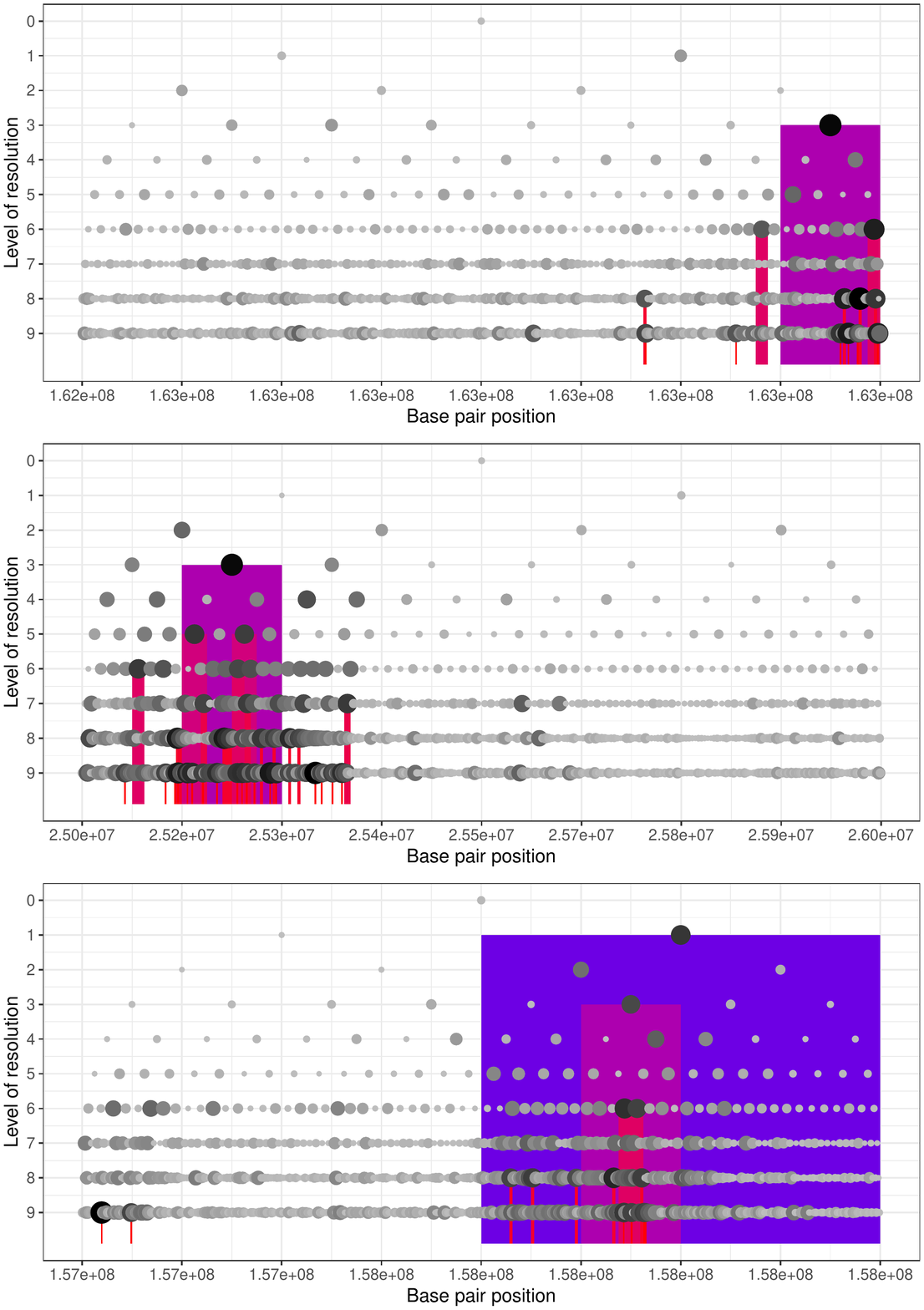}
\end{center}
\caption{Loci discovered by Wavelet Screaming, ordered by increasing p-value (for $d$ coefficients). Top panel: $p=3\times 10^{-12}$, downstream of \textit{EBF1}. Middle panel: $p=2.6 \times 10^{-6} $, previously unreported loci for gestational duration. Lower panel: $p=3.4 \times 10^{-6}$, upstream of \textit{EBF1}.}
  \label{fig:loci2}
\end{figure}

We used this distribution to compute the p-values for each locus. We identified three loci for the $d$ coefficient and one locus for the $c$ coefficient with p-values below $\frac{0.05}{6000}\approx 8.3 \times 10^{-6}$. The discovered loci are depicted in Figure \ref{fig:loci2}. The first and third loci surround \textit{EBF1}; notably, the main SNP from the published meta-analysis is located in the right part of the third plot of Figure \ref{fig:loci2}. In addition, the second locus in Figure \ref{fig:loci2} is located in a regulatory region containing a promoter and multiple transcription factors. Furthermore, this locus is located less than 1 Mb from two genes, which suggest that it might be involved in their regulation. This locus therefore warrants further investigation. 

We use the classic pyramidal wavelet decomposition representation to display the Bayes Factors corresponding to each wavelet coefficient, with point size and darkness representing their values (i.e., the highest Bayes factors are marked by the darkest and largest points). Furthermore, if a Bayes factor is larger than one (i.e., a region contributing to $\hat{\Lambda}$), we highlight the region corresponding to the wavelet coefficient.

\section{Discussion}

In this paper, we introduce Wavelet Screaming as a novel and more powerful alternative to the classic GWAS methodology. It offers a more flexible modeling scheme than the standard single-point testing approach and substantially improves the discovery rate. 

In future development, we aim to expand this tool to include phenotypes on non-ordered scales (e.g., blood types or psychiatric phenotypes), which are normally treated in a case-control manner and not by multinomial regression because of computational burden and power. By exploiting reverse regression, we can include such phenotypes in the predictor matrix by coding them in a similar way to ANOVA. In addition, by exploiting reverse regression, we can also easily adapt this method to the multiple-phenotype GWAS setting \citep{MULTPh}.

This development is made simpler by the results of \cite{ZG17}, in which the authors showed that the parameter of the Bayes factors law depends primarily on the singular values of the matrix of regression and the number of parameters tested. As the regression matrix remains constant across all loci, location and scale, we can compute these parameters only once, enabling a fast computation of p-values. This makes Wavelet Screaming a suitable method to study new phenotypes that are not easily handled by the mainstream GWAS method.

In case of small sample sizes ($n < 1000$), additional parameter computations are needed. Nevertheless, it is still possible to run an efficient screening using a two-stage procedure: first, using the (non-conservative) asymptotic p-value approximation with the non-centrality parameter set as zero, one may select loci that pass the desired threshold. Second, for the selected loci, one may compute the non-centrality parameter of each regression and then compute the specific distribution for these loci (correct p-value).

By contrast, when the sample sizes are large ($n >10^6$, like in \cite{BigWAS}), the user will face the well-known Bartlett's paradox \cite{BP}, which implies that the Bayes factors would converge to $0$. Despite the conservativeness that the paradox brings to discovery, it can also annihilate the discovery capacity. In this setting, the choice of $\sigma_b$ can be crucial. \cite{SS07} suggest to average through some different values of $\sigma_b$ to obtain the proper Bayes factor, which is unrealistic in our setting. In our future work, we will be investigating how to attain a more optimal value for $\sigma_b$ in term of the discovery capacity.

Due to the complexity of the test statistics, it is hard to infer directly how power would be influenced by the parameters. Future work should focus on exploring the power behavior under different conditions (e.g., sample size, variance explained, unequally distributed effect between SNPs, etc.).

Wavelet Screaming is similar to the``Gene- or Region-Based Aggregation Tests of Multiple Variants" described by \citep{LG14} and others (often referred to as burden tests). In these methods, the genetic variants are summed up in a given region to construct a genetic score that are subsequently used in the regression. To some extent, our approach is analogous to extending/generalizing this by performing a multi-scale regional test. Indeed, by using the $c$ coefficients of the wavelet decomposition instead of the $d$ coefficients, we essentially perform a multi-scale burden test. As shown in the simulation, the $c$ coefficient works well when the SNP alleles have an effect in the same direction, which is the main assumption of the burden test. However, our method is the first region-based method that can confidently be applied using $d$ coefficients, without the need to worry much about errors in the coding itself. Indeed, all the current regional methods assume a mono-directional effect setup.

Lastly, our methodology is highly versatile in its applicability to various "omics" data. We intend to investigate its application to, e.g., methylation data, and the feasibility of adding one more level of hierarchy to extend it to multi-omics analyses. 

\section{Software}
The Wavelet Screaming method is distributed as an \textit{\textbf{R}} package. In addition to the analysis code, the package contains a data visualization tool to help elucidate the underlying mechanisms detected by Wavelet Screaming. Our Wavelet Screaming package is available at https://github.com/william-denault/WaveletScreaming. To perform an analysis, the user only needs to specify one parameter ($\sigma_b$). We provide a detailed example of how to use our package in the help function \textit{wavelet\_screaming} based on simulated data. We show how to compute $\lambda_1$ from our package and how to simulate  $\hat{\Lambda}$ under the null. Finally, the user can visualize the output of the \textit{wavelet\_screaming} as depicted in \ref{fig:loci2}  using the  \textit{plot\_WS} function.

\section*{Acknowledgements}
This project was funded by the Research Council of Norway (RCN) from two project: Genetic Factors in Timing of Birth, grant 249779, and in part by the RCN's Centres of Excellence funding scheme, grant 262700. The funders had no role in study design, data collection and analysis, decision to publish, or preparation of the manuscript. We are also grateful to Jonas Bacelis for his detailed comments on the paper, and to Gatien Ricotier for his thorough review of the package.

\section{Citations and References}
 
\bibliographystyle{rss}
\bibliography{example}

\appendix

\section{Additional Figure}
\begin{figure}[!ht]
\begin{center}
\includegraphics[scale=0.4]{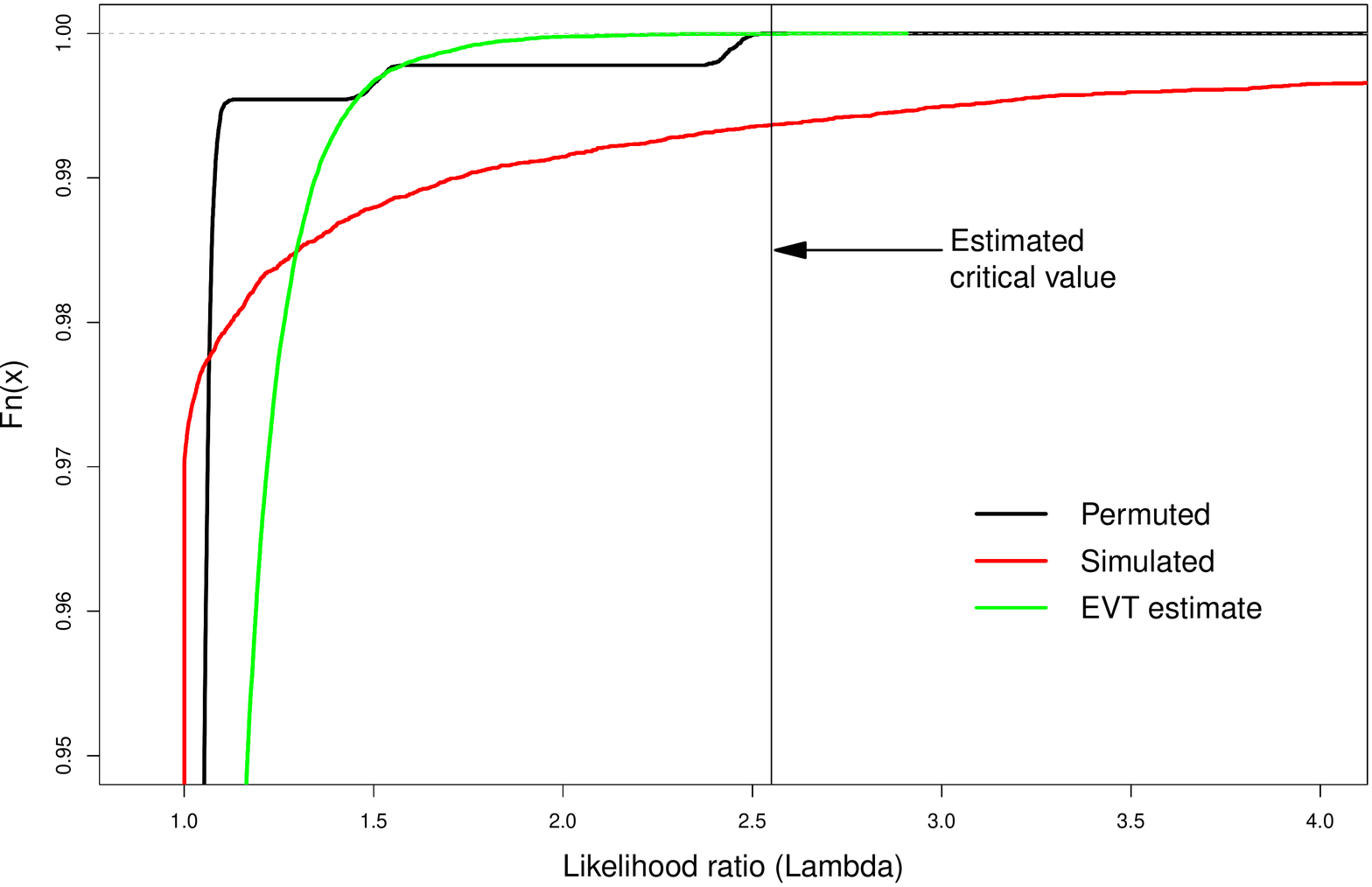}
\caption{A zoomed in version of the upper left part of Figure 3.}

  \label{fig:ECDFb}
  \end{center}
\end{figure}
\end{document}